**Title: Characterizing Secondary Neutrons at BLIP for Isotope Production Applications**

**Authors: Wilson Lin[a*], Michael A. Skulski[a,1], Cathy S. Cutler[a], Dmitri G. Medvedev[a], Jonathan T. Morrell[a*]**

[a]Brookhaven National Laboratory, Upton, New York 11973, USA

*Corresponding authors, wlin4@bnl.gov, jmorrell@bnl.gov

[1]Present address at Northstar Medical Radioisotopes LLC, Beloit, Wisconsin 53511, USA

**Abstract:**

Fast secondary neutrons created at the Brookhaven Linac Isotope Producer (BLIP) facility following proton irradiation were characterized by the foil activation technique and compared with FLUKA Monte Carlo simulations. The FLUKA-simulated neutron flux was spectrally adjusted following the maximum entropy formalism using the International Reactor Dosimetry and Fusion File (IRDFF-II), with predictions agreeing with experimental measurements to within 9% following the adjustment procedure. A multitude of degrader configurations were simulated to assess the feasibility of improving the fast ($E_n$ > 20 MeV) secondary neutron yield at the proposed neutron target position ("N-slot"). A configuration where the N-slot is closest to the proton degrader produced the highest fast neutron yield, with tungsten degraders achieving the best performance. Assuming the optimized target-degrader configuration proposed in this work, we discuss potential isotope production opportunities with secondary neutrons. In most cases the yields are in the order of several mCi.
**Keywords:** secondary neutrons, spectral adjustment, isotope production, BLIP, FLUKA

**Highlights**

- FLUKA simulated secondary neutrons generally agreed with foil activation measurements
- Tungsten degraders may increase fast secondary neutrons >3x compared to current setup at BLIP
- BLIP's N-slot may be used as an additional route for isotope production

# 1 Introduction

The Brookhaven Linac Isotope Producer (BLIP) facility [1–3] at Brookhaven National Laboratory (BNL) supplies the United States with isotopes relevant for nuclear medicine, fundamental sciences, and national security [4–7]. Co-produced fast secondary neutrons at BLIP during proton irradiation could be used for generating radionuclides that may be difficult to produce from charged particle reactions or low energy neutrons [8–11]. While low energy neutrons react with nuclei through radiative capture or fission, fast secondary neutrons possess sufficient energy to eject charged nucleons which can yield exotic neutron-rich radionuclides [9,10]. While large thermal neutron fluxes (>$10^{15}$ n/cm$^2$-s) can be achieved with research nuclear reactors, neutrons above 10 MeV are suppressed by several orders of magnitude and no neutrons above 14 MeV are available [12]. Furthermore, since fast secondary neutrons do not generate appreciable heat upon travelling through matter, in contrast to charged particles, the radionuclidic yield can be scaled with increased target mass without concern for increased thermal loading. These considerations, along with limited information on the secondary neutron flux, served as motivation to investigate the utility of the secondary neutrons at BLIP (referred to as "N-slot production"). Results from this work will help establish future N-slot applications.

Secondary neutrons arising from 100 MeV incident protons at the Isotope Production Facility (IPF) in Los Alamos National Laboratory (LANL) had been previously characterized by MCNP simulations and spectrally adjusted using select activation products in monitor foils [8,13]. These experiments produced fast secondary neutrons resulting from proton irradiation of Rb and Ga targets. Although most of the produced radionuclides in their work had activities ~1 µCi, scaling by longer irradiation times (until saturation) and increasing target masses will improve yield by several orders of magnitude. It is anticipated that the secondary neutron spectra at BLIP will differ from that of IPF due to the higher proton energy, though differences in irradiation geometry and degrader materials will also change the neutron spectra.

In this manuscript, the secondary neutron flux at BLIP produced during routine isotope production was indirectly characterized by using experimentally measured activity from monitor foils to spectrally adjust FLUKA [14,15] simulated neutron spectra. FLUKA was chosen for this application because of its widespread use among high energy accelerator facilities, direct activity simulation and generally good agreement with experimental data [16,17]. Due to similarities between the irradiation facility at IPF and BLIP, similar procedures [8,13] were adopted in this work. Different energy degrader materials and target array geometries were then simulated in FLUKA to evaluate improvements in the fast neutron flux. The manuscript concludes by discussing potential applications of BLIP's N-slot, with emphasis on producing isotopes for medical applications.

# 2 Methods

## 2.1 Characterizing neutron flux at the N-slot

### 2.1.1 Experimental BLIP target array

Parameters of the BLIP target array along with beamline components are provided in the supplemental information. The target array begins with an isotope production target(s) (IPT), followed by a Cu vacuum degrader, two Al degraders and a Cu beam stop. The "N-slot" where neutron target(s) can be positioned is located at the end of the target array. In between each of the degrader/targets, there is a water-cooling channel. For the experiment, the N-slot was occupied by an Al holder with monitor foils arranged in four quadrants to assess potential asymmetry in the neutron flux. Four Bi foils were placed into each of the quadrants and Al, Ni, Zn, Co, Y, Au were placed behind the Bi foils (facing the positive beam axis, top left: Al and Y; top right: Ni and Au; bottom left:

Co and Zn. The irradiation setup is shown in Figure 1). Nominally, the foils were 25 mm x 25 mm with thicknesses 0.25 mm for Bi, Al, Ni and Zn, 0.127 mm for Y and 0.10 mm for Au and Co, with each being >99.95% pure (foils purchased from Goodfellow). All foils were encapsulated in Kapton tape (0.025 mm film, 0.035 mm Si adhesive) while the holder itself was sealed with an aluminum lid secured using Allen screws at each corner. A Viton gasket was used to prevent water leakage. The reported foil masses have an uncertainty of ±0.5 mg (see Table 1).

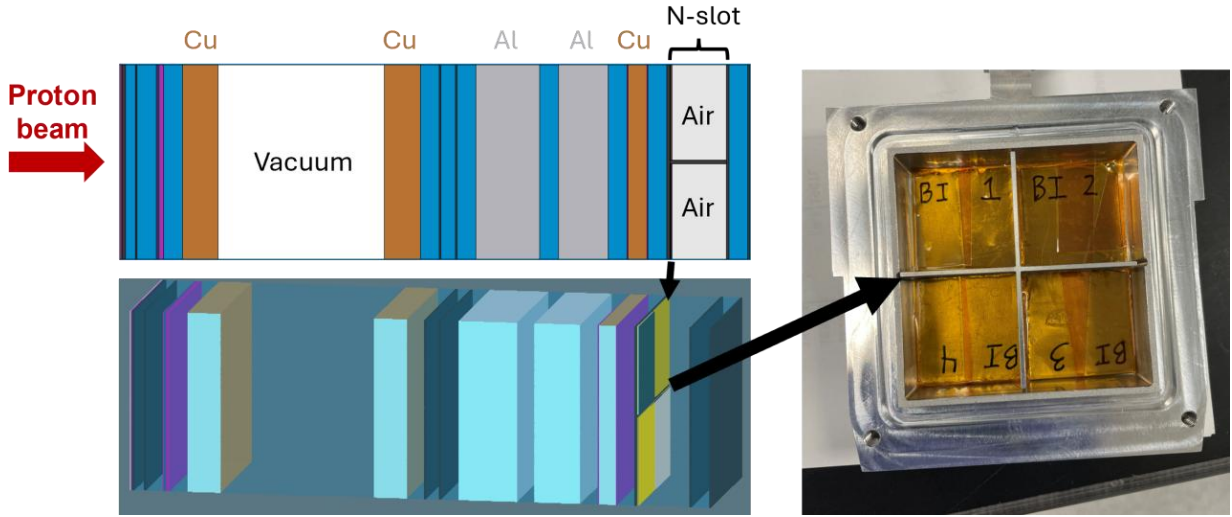

*Figure 1 The irradiation setup for characterizing secondary neutrons during irradiation of isotope production targets. (Left, top) The target array consists first of an IPT, then Cu and Al degraders until the proton energy reaches ~35 MeV (blue coloring represents water channels). The Cu beam stop after the Al degraders prevents primary protons from reaching the N-slot. (Left, bottom) A 3D schematic to visualize foil placements at the N-slot, where four quadrants were sectioned off for evaluating beam asymmetry. (Right) A physical image showing the front of the N-slot viewed axially, along the positive beam axis. Foils were held vertically by custom stainless-steel springs (not shown).*

*Table 1 Foil masses along with FLUKA simulated and experimentally measured activities for observed radionuclides. Reported activities are decay corrected to end of bombardment without subtracting contributions from parent radionuclide(s). Since the mass normalized activity of the four Bi foils agreed within uncertainty (<5%) for both simulations and experiments, the reported values are unweighted averages of all four foils. Monitor products used in the spectral adjustment process are indicated by \*, where some IRDFF reactions were unused due to suspected contributions unrelated to neutron activation. "FLUKA-only" refers to activities derived directly from FLUKA simulations and "IRDFF-II" refers to activity calculated from folding the FLUKA simulated neutron flux with IRDFF-II [18] cross sections. "N/A" refers to "not applicable".*

| Monitor foil | Mass (g) | Monitor product | Exp. mass normalized activity (μCi/g) | FLUKA-only mass normalized activity (μCi/g) | IRDFF-II mass normalized activity (μCi/g) | FLUKA-only/Exp. | IRDFF-II/Exp. | Adjusted IRDFF-II/Exp. |
|---|---|---|---|---|---|---|---|---|
| Al | 0.4281 | $^{24}$Na* | 340±20 | 300±10 | 440 | 0.88±0.06 | 1.3 | 1.1 |
|  |  | $^{22}$Na | 0.076±0.007 | 0.060±0.005 | N/A | 0.79±0.10 | - | - |
| Au | 1.144 | $^{198}$Au | (1.56±0.09) x10$^4$ | (1.3±0.3) x10$^4$ | N/A | 0.83±0.20 | - | - |
|  |  | $^{196}$Au* | 65±4 | 67±1 | 76 | 1.0±0.1 | 1.2 | 1.0 |
| Bi | 1: 1.5963 | $^{207}$Bi* | 0.030±0.003 | 0.0282±0.0005 | 0.028 | 0.94±0.10 | 0.93 | 0.95 |
| | | 2: 1.6342 | $^{206}$Bi* | 36±2 | 37.9±0.8 | 33 | 1.1±0.1 | 0.92 | 0.95 |
| | | 3: 1.6498 | $^{205}$Bi* | 11.7±0.7 | 13.8±0.3 | 10.9 | 1.18±0.08 | 0.93 | 0.94 |
| | | 4: 1.6695 | $^{204}$Bi* | 220±10 | 278±8 | 280 | 1.3±0.1 | 1.3 | 1.1 |
| | | | $^{203}$Bi | 150±9 | 193±6 | N/A | 1.3±0.1 | - | - |
| | | | $^{202}$Tl | 0.28±0.02 | 0.26±0.07 | N/A | 0.93±26 | - | - |
| Co | 0.5774 | | $^{60(m+g)}$Co | 28±2 | 43±2 | N/A | 1.5±0.1 | - | - |
| | | | $^{58(m+g)}$Co | 10.3±0.6 | 9.6±0.2 | 9.4 | 0.93±0.06 | 0.91 | 0.85 |
| | | | $^{57}$Co | 1.04±0.06 | 1.12±0.04 | 0.93 | 1.08±0.07 | 0.89 | 0.92 |
| | | | $^{56}$Co | 0.53±0.03 | 0.70±0.03 | N/A | 1.3±0.1 | - | - |
| | | | $^{59}$Fe* | 1.33±0.08 | 1.37±0.08 | 1.4 | 1.03±0.09 | 1.1 | 0.98 |
| | | | $^{56}$Mn | 400±30 | 320±20 | 380 | 0.80±0.08 | 0.95 | 0.85 |
| | | | $^{54}$Mn | 0.28±0.02 | 0.26±0.02 | N/A | 0.93±0.10 | - | - |
| | | | $^{52(m+g)}$Mn | 1.43±0.09 | 0.93±0.09 | N/A | 0.65±0.08 | - | - |
| Ni | 1.4721 | | $^{57}$Ni | 50±3 | 54.6±2 | 36 | 1.1±0.1 | 0.72 | 0.71 |
| | | | $^{56}$Ni | 0.55±0.03 | 1.25±0.15 | N/A | 2.3±0.3 | - | - |
| | | | $^{60(m+g)}$Co | 0.025±0.002 | 0.021±0.001 | 0.021 | 0.84±0.08 | 0.84 | 0.77 |
| | | | $^{58(m+g)}$Co | 8.5±0.6 | 10.1±0.2 | 9.8 | 1.2±0.1 | 1.2 | 1.1 |
| | | | $^{57}$Co | 1.9±0.1 | 2.62±0.02 | N/A | 1.4±0.1 | - | - |
| | | | $^{56}$Co | 2.5±0.2 | 2.67±0.04 | N/A | 1.1±0.1 | - | - |
| | | | $^{59}$Fe | 0.067±0.005 | 0.031±0.008 | N/A | 0.46±0.12 | - | - |
| | | | $^{54}$Mn | 0.18±0.01 | 0.180±0.007 | N/A | 1.0±0.1 | - | - |
| | | | $^{52(m+g)}$Mn | 3.6±0.2 | 3.1±0.2 | N/A | 0.86±0.07 | - | - |
| Y | 0.3544 | | $^{88}$Y | 6.5±0.4 | 6.8±0.2 | 5.9 | 1.0±0.1 | 0.91 | 0.85 |
| | | | $^{87(m+g)}$Y | 138±8 | 120±4 | N/A | 0.870±0.058 | - | - |
| | | | $^{86}$Y | 220±10 | 240±10 | N/A | 1.1±0.1 | - | - |
| | | | $^{84}$Rb | 0.88±0.05 | 0.41±0.08 | N/A | 0.47±0.10 | - | - |
| | | | $^{83}$Rb | 0.59±0.04 | 0.57±0.09 | N/A | 0.97±0.17 | - | - |
| Zn | 1.1252 | | $^{67}$Cu | 12.4±0.7 | 11.5±0.7 | 7 | 0.93±0.08 | 0.56 | 0.55 |

The target array was irradiated at 150 µA for 30 minutes with Gaussian beam rastered at d=11.5 mm and 5.5 mm circles, 1:10 respectively, with incident proton energy of 200 MeV. After irradiation, the holder was opened in a hot cell by removing the screws and the foils were carefully extracted from the hot cell for non-destructive gamma spectroscopy using High Purity Germanium (HPGe) detectors. The foils were first counted approximately 10 h post end of bombardment (EOB) then periodically counted thereafter to reduce uncertainties for longer-lived radionuclides. The detectors were calibrated with sources traceable to National Institute of Standards and Technology ($^{22}$Na, $^{241}$Am, $^{152}$Eu, $^{137}$Cs, $^{57}$Co, $^{54}$Mn, $^{60}$Co and $^{109}$Cd, nominally 1 µCi each with ±3% uncertainty in reported activity). Photon peak fitting and detector efficiency calibration for residual activity quantification were performed using the CURIE Python library [19]. Peaks with >10% (combined systematic and statistical) uncertainty were discarded, and final decay corrected EOB activities were calculated based on the weighted average of all valid peaks, weighted by inverse relative variance. Contributions from parent radionuclide(s) were not subtracted from the activity of the daughter to simplify analysis when comparing with simulations. The uncertainty in efficiency calibration was set at 5% by inspection. A sample HPGe calibration spectrum and resulting efficiency calibration is shown in Figure S1. Final uncertainties were obtained by propagating independent error terms quadratically. Uncertainties in this work are reported as one standard deviation about the mean. The gamma emissions used in this work are provided in Table S1. Unless otherwise stated, percent deviation refers to 100%x(10$^{(abs(log10(Ratio)))}$ - 1), where "Ratio" is $A_{calc}/A_{meas}$.

### 2.1.2 FLUKA simulated experimental target array

The target array described above was reproduced in FLUKA (version 4-4.1) [14,15] using the FLAIR GUI [20]. Incident protons (primary particles) were sampled from a uniform disk with radius 1.6 cm to approximate the average behavior of the irradiation at BLIP based on the current density [2]. The following physics models were used: coalescence, the new evaporation model with heavy fragment evaporation, radioactive decay, the default library for low energy neutrons <20 MeV, and the PEANUT package with full ion transport [16,21]. FLUKA simulations were performed using 6x10$^8$ primary particles except when predicting monitor foil activities (6x10$^9$ primary particles). A custom Fortran code was included to discriminate neutrons in different energy bins for visualizing the neutron flux map. Activities obtained directly from FLUKA simulations are referred to as "FLUKA-only" to distinguish between activities derived from folding FLUKA simulated neutron spectra with reaction cross sections.

As previously mentioned, reported activities in this work include decay correction to EOB and decay of shorter-lived parent radionuclide(s). For example, if 100 atoms of $^{52m}$Mn and 300 atoms of $^{52g}$Mn were predicted at EOB from FLUKA, then the activity would be calculated using (100*BR$_{52gMn}$ +300) atoms and multiplying by the decay constant with EOB decay correction, where BR$_{52gMn}$ refers to the branching ratio of the isomer to the ground state. This assumption is valid for spectra acquired >4 half-lives of the longest-lived parent and approximates the following equation ( $\lambda_p \gg \lambda_d$ and $\lambda_p t \gg 1$):

$$N_d(t) = N_p(0)\frac{\lambda_p BR_{p \to d}}{\lambda_d - \lambda_p}\left(e^{-\lambda_p t} - e^{-\lambda_d t}\right) + N_d(0)e^{-\lambda_d t} \approx e^{-\lambda_d t}\left[N_p(0)BR_{p \to d} + N_d(0)\right] \quad (1)$$

### 2.1.3 Spectral adjustment of secondary neutron flux

Monitor reactions were taken from the IRDFF-II library [18] to compare FLUKA simulated neutron flux with experimentally measured activity due to its rigorous evaluation. IRDFF-II cross sections are shown in Figure 2, which includes all residuals that could be produced from the foils used in this work (some residuals were not observed).

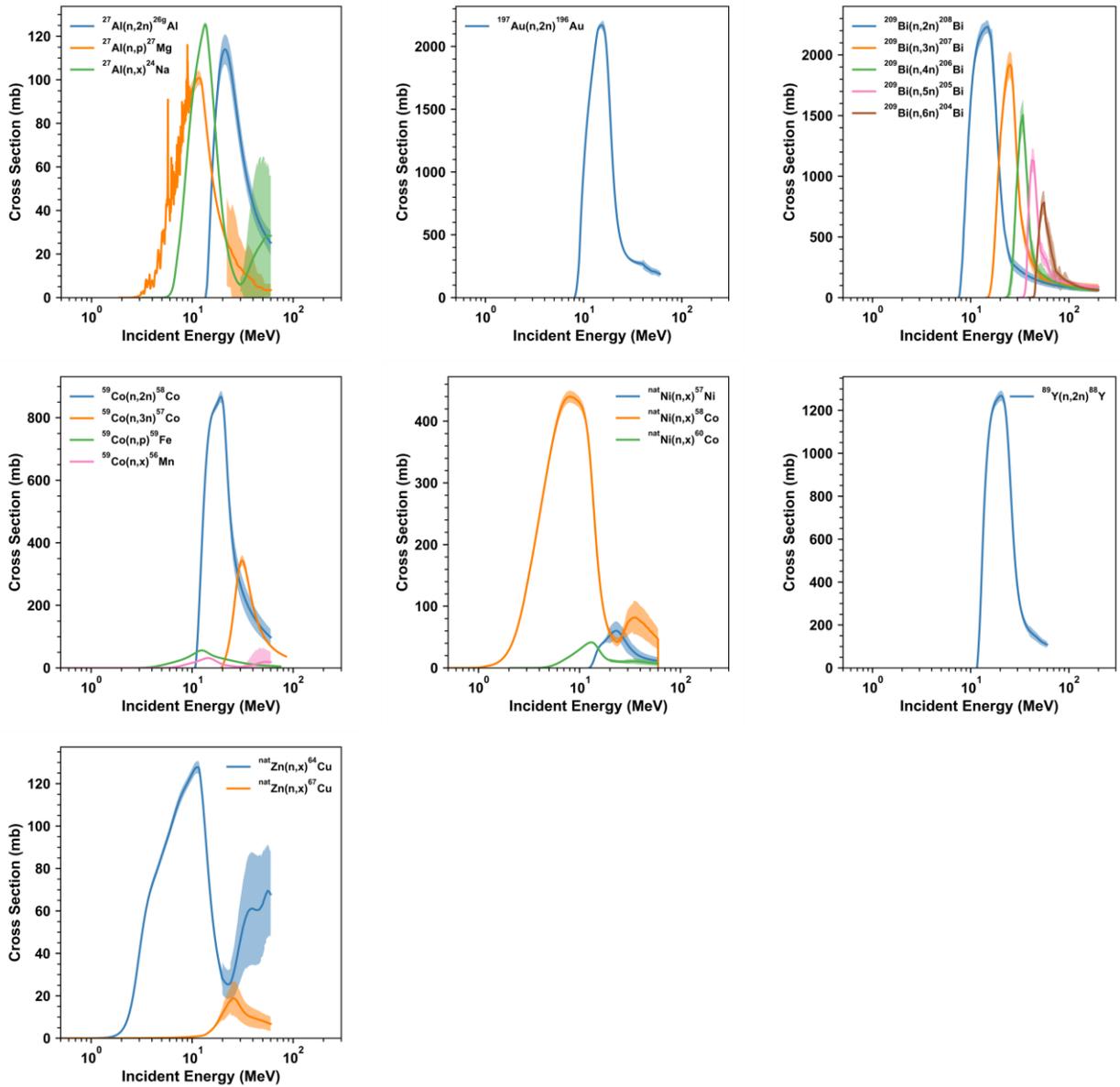

Figure 2 IRDFF-II cross sections [18] relevant for each monitor foil used in this work. The dark shading indicates the uncertainty of one standard deviation about the mean.

Spectral adjustment was performed using a homemade Python program similar to the MAXED code written in FORTRAN [22]. The Python software was created due to the inability to alter parameters in existing spectrum adjustment codes, such as SAND-II, and to facilitate future adaptations. The maximum entropy method was chosen because of its least biased approach and potential for uncertainty propagation [22,23]. Briefly, the measured activities were decay corrected and transformed into the number of produced nuclides per target nuclide. The MAXED code maximizes the cross entropy $S$ [22,23]:

$$S = -\sum_i \phi_i \ln\left(\frac{\phi_i}{\phi_i^{Def}}\right) + \phi_i^{Def} - \phi_i \quad (2)$$

Subject to the following set of constraints:

$$N_k + \epsilon_k = \sum_i \sigma_{ki}\phi_i, \quad k = 1,..m \quad (3)$$

$$\sum_k \frac{\epsilon_k^2}{\delta_k^2} = \chi^2 \quad (4)$$

Where $N_k$ is the number of produced nuclides per target atom for reaction channel $k$, $\epsilon_k$ is the unknown error in $N_k$, $\sigma_{ki}$ is the cross section for the reaction channel at energy $i$, $\phi_i$ is the neutron fluence (Def, short for default, refers to the unadjusted fluence), $\delta_k$ is the uncertainty and $\chi^2$ is the chi-square statistic which is set as $m$, the total number of reaction channels. Using the method of Lagrange multipliers, maximizing the following equation becomes akin to maximizing the constrained cross entropy:

$$Z = -\sum_i \phi_i^{Def} e^{-\sum_k \lambda_k \sigma_{ki}} - \sqrt{\chi^2 \sum_k (\lambda_k \delta_k)^2} - \sum_k N_k \lambda_k \quad (5)$$

Where, in this equation, $\lambda_k$ is the Lagrange multiplier for reaction channel $k$. Once the potential function is maximized, the adjusted spectrum is calculated as:

$$\phi_i = \phi_i^{Def} e^{-\sum_k \lambda_k \sigma_{ki}} \quad (6)$$

In this work, the Python program minimizes -$Z$ which is identical to maximizing $Z$.

## 2.2 Assessing the impact of degrader materials on neutron yield for isotope production applications

Various target arrays were simulated using FLUKA to evaluate the production capacity of the N-slot. Since different degrader materials have varying densities and thermal load constraints, the thicknesses of the degraders were chosen such that ~5 kW thermal burden for each degrader was not exceeded. Tungsten was chosen due to its higher density which could position the N-slot closer to the initial source(s) of fast neutrons. Beryllium was selected based on potentially improved fast neutron yield (lighter nuclei), though the material had to be a compound with higher density such as BeO to satisfy the geometrical constraints. Finally, Cu and Al were chosen based on their good thermal conductivity, low activation and a long history of successful use at BLIP. In all cases, the same IPT was placed at the front of the array and 5 mm of Sc was used as the beam stop. The Sc beam stop was chosen as an additional target for isotope production. As a practical method for evaluating yield from fast secondary neutrons, a simulated 1 g pellet of $^{226}$RaCl$_2$ salt with diameter 1 cm (density 4.9 g/cm$^3$) was placed at the center of the N-slot. This setup provides an estimate of $^{225}$Ra production through $^{226}$Ra(n,2n)$^{225}$Ra ($E_{th}$ ~7 MeV) for potential use to generate $^{225}$Ac which is an important radiotherapeutic. All mass normalized activities refer to the activity divided by the target mass and should not be confused with specific activity.

# 3 Results

## 3.1 Characterizing secondary neutrons at the N-slot from isotope production targets

FLUKA simulations were first performed to predict the activities of various radionuclides for experimental planning. The FLUKA calculated neutron flux is shown in Figure 3, where the neutron distribution appears to be axially symmetric. Following a simulation with 6x10$^9$ primary particles at 200 MeV, many of the desired radionuclides from monitor reactions [13,18] were observed with <5% uncertainty (see Table 1). Likely due to water moderation, FLUKA predicts some thermal neutrons at the N-slot. However, the secondary thermal neutron flux is several orders of magnitude lower than common research reactors [12]. Proton induced reactions were also found at the N-slot, where additional protons could have been created from fast secondary neutron interactions. To test this hypothesis, the density of the Cu beam stop was increased to a super-physical value of 45.5 g/cm$^3$ to ensure complete removal of primary protons. The results confirmed that protons at the N-slot are likely secondary/tertiary in nature and not from poor energy degradation (see Figure S2).

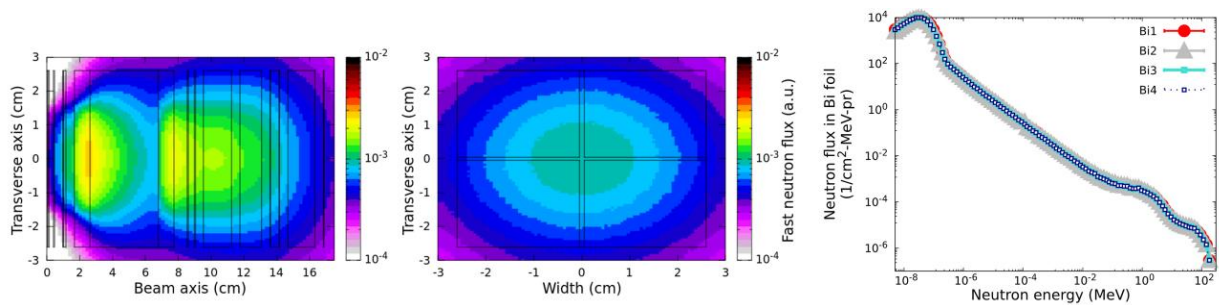

*Figure 3 The neutron flux map of the entire target array (left) and at the N-slot (middle) for $E_n$ >20 MeV. In addition to the X-Y neutron flux map, the differential neutron flux in each Bi monitor foil (right) also suggests minimal beam asymmetry.*

The experimentally measured activities are reported in Table 1, where comparisons for FLUKA-only and FLUKA+IRDFF-II predicted yields are also provided. In general, both FLUKA-only and FLUKA+IRDFF-II predictions agree well with measured activities. However, FLUKA+IRDFF-II typically underestimated the radionuclidic yield compared to

FLUKA derived activities, potentially due to the presence of secondary protons. Major deviations (≥20%) between FLUKA-only predictions and experimental measurements for monitor reactions were $^{209}$Bi(n,x)$^{204}$Bi and $^{59}$Co(n,x)$^{56}$Mn with predicted $A_{FLUKA}/A_{Exp}$ of 1.3 and 0.80, respectively. Other reactions, i.e., not present in IRDFF-II, from fast neutron activation with observed deviations ≥20% were $^{nat}$Ni(n,x)$^{56}$Ni, $^{nat}$Ni(n,x)$^{59}$Fe, $^{89}$Y(n,x)$^{84}$Rb, $^{59}$Co(n,x)$^{52}$Mn, $^{nat}$Ni(n,x)$^{57}$Co, $^{209}$Bi(n,x)$^{203}$Bi, $^{59}$Co(n,x)$^{56}$Co and $^{27}$Al(n,x)$^{22}$Na with predicted ratios of 2.3, 0.46, 0.47, 0.65, 1.4, 1.3, 1.3 and 0.79, respectively. Two thermal neutron capture products, $^{198}$Au and $^{60}$Co, were observed as well, though FLUKA-only overestimated the yield for $^{59}$Co(n,γ)$^{60}$Co by 50%.

Nevertheless, with careful selection of the monitor reactions, the FLUKA predicted neutron spectrum was adjusted following the maximum entropy formalism. Additional information related to the spectral adjustment process is provided in Figure S3. Since FLUKA+IRDFF-II activities were already in good agreement with experimental data prior to the adjustment procedure, only minor differences exist between the adjusted and original neutron flux (Figure 4). Still, the adjusted spectrum predicts yields that align better with experimental measurements, with all monitor reactions used in the adjustment process having percent differences within ±9%.

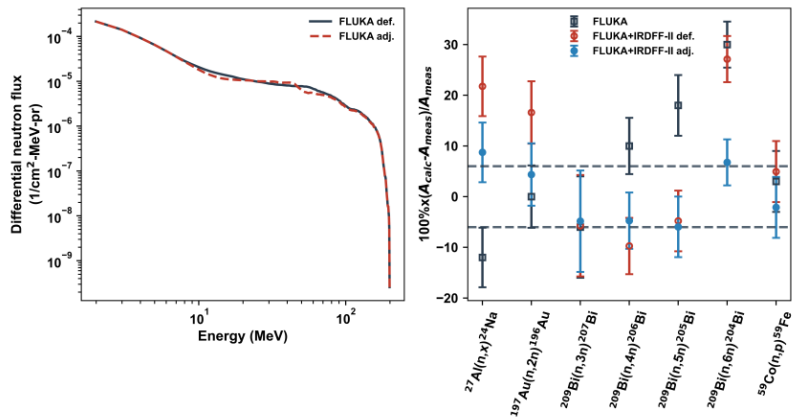

*Figure 4 The FLUKA predicted neutron spectrum before and after spectral adjustment (left). (Right) The percent differences between measured and calculated activities for each of the monitor reactions used in the spectral adjustment process. Eye guides at ±6% refer to the minimum uncertainty from the measurements. All error bars correspond to the percentage uncertainty in the measurement and were not propagated.*

## 3.2 Improving fast secondary neutron yield for isotope production applications

After characterizing the neutron flux at the N-slot, different target arrays were investigated to identify factors that can improve the fast secondary neutron yield for isotope production efforts. The neutron flux maps for each target array investigated are shown in Figure 5 and the relative $^{225}$Ra yields from the $^{226}$RaCl$_2$ target are provided in Table 2. A cut-off $E_n$>20 MeV in Figure 5 was chosen to best represent fast secondary neutrons useful for producing exotic nuclides with high energy thresholds. For reference, the simulated saturation yield of $^{225}$Ra at EOB from FLUKA was ~0.2 mCi/μA-g (~7 MBq/μA-g, saturation activity normalized by $^{226}$RaCl$_2$ mass and proton current) from the isotope production array.

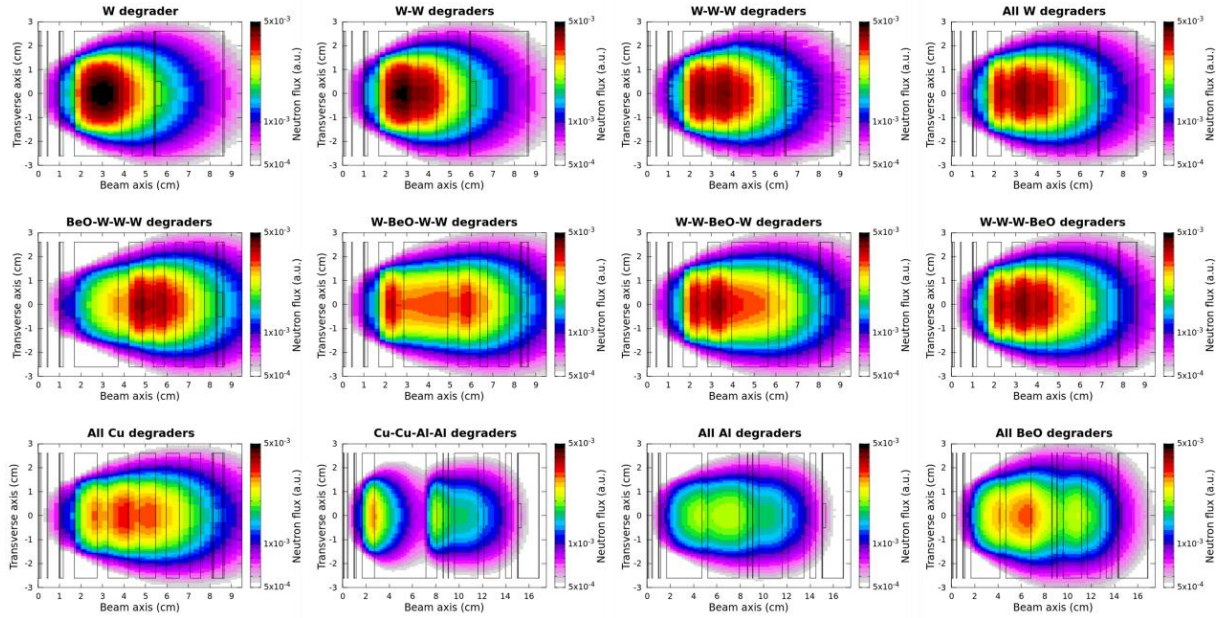

Figure 5 FLUKA simulated neutron flux maps ($E_n$>20 MeV) for each of the different target arrays investigated in this work. The geometrical constraints were based on thermal and physical considerations and chosen to match as closely to the original setup as possible.

Table 2. The $^{225}$Ra yield normalized relative to the isotope production array (IPT, Cu-Cu-Al-Al) from each of the target arrays investigated in this work. The isotope production array can produce ~0.2 mCi/μA-g $^{225}$Ra at the N-slot. All target arrays included the same IPT as the BLIP target array in the initial position and Sc beam stop before the N-slot. Except for the three larger W degraders, all other setups had 4 energy degraders. Total relative uncertainty for all target arrays was <10%.

| Degrader array   | $^{225}$Ra yield relative to Cu-Cu-Al-Al |
|------------------|------------------------------------------|
| W                | 3.1                                      |
| W-W              | 2.8                                      |
| BeO-W-W-W        | 2.4                                      |
| W-W-W            | 2.0                                      |
| W-BeO-W-W        | 1.8                                      |
| W-W-BeO-W        | 1.5                                      |
| W-W-W-BeO        | 1.6                                      |
| W-W-W-W          | 1.8                                      |
| Cu-Cu-Cu-Cu      | 1.7                                      |
| BeO-BeO-BeO-BeO  | 1.4                                      |
| Al-Al-Al-Al      | 0.80                                     |

The neutron flux maps show that fast secondary neutrons ($E_n$>20 MeV) preferentially scatter in the forward direction with substantial intensity attenuation with distance. Thus, the best performing array in the first set of comparisons included four W degraders, resulting in the shortest distance to the N-slot. Also, the increase in relative yield for the BeO-BeO-BeO-BeO stack over Cu-Cu-Al-Al suggests that a combination of W and BeO could potentially improve the neutron yield further if the target array can be constrained to one enclosure.

A second set of simulations investigated the positioning of BeO with three W degraders and varying the number of W degraders. Although there was slight improvement by setting the first degrader to be BeO (remaining three W) compared to only using W degraders, the suboptimal thermal and physical characteristics of Be related materials may limit this design implementation. Alternatively, having fewer W degraders to reduce the distance to the N-slot resulted in marked improvements in $^{225}$Ra yield (>3x relative to the isotope production array).

# 4 Discussion

## 4.1 Experimental validation of FLUKA for characterizing fast secondary neutrons at the N-slot

FLUKA simulations agreed well with experimentally measured activities for the evaluated monitor reactions. The activity yields (normalized by mass, current and irradiation time) in this work relative to those of IPF [8] are approximately 2x greater for activation products from monitor reactions, likely due to the increased incident proton energy (see Table S3). Short-lived radionuclides such as $^{27}$Mg were not observed and/or had low counting statistics due to logistical challenges and dose constraints in removing the foils from the BLIP hot cell. Although some of these reactions would have been optimal for spectral adjustment due to the lack of available proton channel(s) to produce the same product, observed radionuclides in this work still span most of the fast neutron energy range. It may also be useful for future applications to surround monitor foils/targets with neutron absorbers [24] such as Cd to reduce unwanted activation from thermal neutrons. Negligible impact to the fast secondary neutron is expected for most neutron absorbers.

Although FLUKA predicted the activity and neutron spectrum with generally good agreement between experimental measurements, several reaction channels still had >30% deviation. For some reaction channels, such as $^{89}$Y(n,x)$^{84}$Rb, this discrepancy could be explained in part by FLUKA's assumption of equally sharing isomeric and ground states during production [17]. Alternatively, the lack of evaluations for many of these reaction channels could also lead to unexpected results, especially when there are secondary protons (and potentially α-particles) with energies above the reaction threshold. These considerations led to additional screening for evaluated monitor reactions in the spectral adjustment process (see Figure S2 for the proton flux distribution). Predicted yields from folding the resulting spectrally adjusted spectrum with IRDFF-II cross sections were within ±9% of the experimentally measured values and had reduced $\chi^2 \sim 1$, which suggests that the spectrum was not over/under-fitted. The deviations from experimental measurements after spectral adjustment in this work are comparable to the data observed at IPF [13], where the researchers used MCNP to predict the neutron flux and the SAND-II code for spectral adjustment. Given that the maximum entropy method results in an analytical solution, future work will attempt to implement the uncertainty propagation as was done in [23], which is not possible with iterative methods such as SAND-II.

## 4.2 Proposed degrader choices and array design for improving fast neutron yield based on FLUKA simulations

After validating FLUKA's utility for predicting fast secondary neutrons, additional sets of simulations aimed to improve the fast neutron yield at the N-slot for potential isotope production applications. The simulations indicated that the best target array design would be to minimize the distance of the N-slot from the degrader(s). This reduction in neutron flux with increased distance is likely due to neutron angular divergence (with some attenuation from neutron-material interactions prior to the N-slot). The attenuation purely from geometry can be observed at the N-slot in Figure 5 for each of the configurations, where the N-slot was modelled with vacuum media. Also, the proton beam spread across each degrader array configuration should have a low impact on the secondary neutron flux at the N-slot since most fast neutrons are produced in regions with little proton beam spread.

From the simulations in this work, W was identified as an ideal degrader material for N-slot applications due to its high density and atomic number. Additionally, W has the highest melting point out of all elements and is much cheaper relative to other potential candidates with high density and atomic number such as Pt and Os. Safety constraints for Os ($OsO_4$) also make Os challenging to work with. The initial set of simulations were performed under the assumption that each energy degrader will survive ~5 kW of thermal load with the given cooling environment. Since W had previously been irradiated under harsher conditions [25–27], the thickness of each W degrader was increased to reduce the number of degraders in the array. However, previous investigations have reported substantial corrosion for W exposed to water under irradiation [25,26] which will necessitate passivation or encapsulating pure W. Tantalum may be considered as an alternative to preclude corrosion, though it has lower density, lower melting point and increased cost. Nevertheless, the FLUKA simulations show that using only one W degrader leads to ~2x $^{225}$Ra yield relative to four equally spaced W degraders and >3x $^{225}$Ra yield relative to the current isotope production array at BLIP.

## 4.3 Potential isotope production using the N-slot at BLIP

Using the optimized neutron yields from just one W degrader, activities for several radionuclides most well-suited for fast secondary neutron production were calculated by folding the neutron spectrum with TENDL cross sections (see Table 3) [28]. Yields calculated from FLUKA+TENDL may not be as precise as FLUKA+IRDFF-II but folding the neutron flux with TENDL is an efficient method for estimating yield due to the large availability of reactions. Using the N-slot, and assuming saturation yield at EOB, ~0.6 mCi/μA-g $^{225}$Ra, 1.4 mCi/μA-g $^{47}$Ca and 0.03 mCi/μA-g $^{195m}$Pt can be produced using the N-slot with $^{226}$Ra, $^{48}$Ca and $^{197}$Au, respectively. Practically, only $^{225}$Ra can be produced in sufficient quantities for clinical applications in this scenario but the ease of targetry could enable routine productions of other radionuclides for preclinical/research applications. Scaling up the mass of more accessible and dense target material such as Au can greatly improve yield without concerns for additional thermal loading. Probing the feasibility of $^{225}$Ra production further, a practical elution time of $^{225}$Ac would be ~15 days which leads to an $^{225}$Ac yield ~30% of the original activity of $^{225}$Ra every elution. This results in a realistic ~0.2 mCi/μA-g $^{225}$Ra for downstream applications, and half of the activity every 15 days thereafter. In comparison with the Joyo fast neutron reactor [29], the N-slot can achieve a relative mass normalized saturation yield of ~10% $^{225}$Ra and >150x reduction in co-produced $^{227}$Ac ($^{227}$Ac can be separated from $^{225}$Ra but still relevant for waste considerations).

For more exotic longer-lived tracer isotopes such as $^{53}$Mn and $^{32}$Si, production pathways are often through heavy ion fusion or high energy spallation, which require dedicated facilities and co-produce relatively large quantities of other radionuclides. Since applications with these long-lived isotopes generally require much less activity than medical applications, fast secondary neutrons can provide a practical alternative. Previous works using charged particle beams produced ~21.5 pg (9.3 fCi) $^{53}$Mn in 6 h ($^{13}$C beam on 9 mg/cm$^2$ $^{46}$Ti) [30] and 100-150 μCi $^{32}$Si in 3 months (proton beam on 100 g KCl, 800 MeV, 1 mA) [31,32]. In comparison, the N-slot can produce similar quantities of $^{53}$Mn with just 0.5 mg $^{54}$Fe in the same irradiation period (150 μA protons) but will require ~10's g $^{34}$S to achieve comparable $^{32}$Si yields. However, in contrast to the spallation production route for $^{32}$Si (which included a 2–4-year decay period prior to processing KCl spallation targets), the N-slot can greatly simplify radiological waste logistics due to the low quantities of co-produced radionuclides.

*Table 3 Proposed radionuclides that could be produced from the N-slot at BLIP which may be difficult to produce otherwise. Calculations were performed by folding the FLUKA predicted neutron spectrum with TENDL cross sections*

[28]. The reported yields are in units of mCi/µA-g, which implies saturation. Entries marked with "*" assume a 2700 h irradiation due to the long half-life.

| Reaction | Half-life | Mass normalized saturation yield at EOB (mCi/µA-g) | Application |
|---|---|---|---|
| $^{226}$Ra(n,2n)$^{225}$Ra | 14.9 d | 0.6 | Parent of $^{225}$Ac used for radiopharmaceutical therapy [33] |
| $^{48}$Ca(n,2n)$^{47}$Ca | 4.536 d | 1.4 | Parent of $^{47}$Sc used for radiopharmaceutical therapy [34] |
| $^{197}$Au(n,x)$^{195m}$Pt | 4.04 d | 0.03 | Radiopharmaceutical therapy [35] |
| $^{54}$Fe(n,x)$^{53}$Mn<br>$^{nat}$Fe(n,x)$^{53}$Mn | 3.7x10$^6$ y | 6.5 x10$^{-8}$*<br>1.7 x10$^{-8}$* | Geochronology tracer [30,36] |
| $^{33}$S(n,x)$^{32}$Si<br>$^{34}$S(n,x)$^{32}$Si<br>$^{36}$S(n,x)$^{32}$Si | 153 y | 1.2 x10$^{-5}$*<br>3.5 x10$^{-5}$*<br>13 x10$^{-5}$* | Oceanography tracer [31,32]. High isotopic purity $^{36}$S is currently unavailable. |

# 5 Conclusion

Foil activation experiments are efficient methods to validate the characterization of fast secondary neutrons from simulations, especially for isotope production facilities such as BLIP and IPF. FLUKA simulated secondary neutron flux at BLIP's N-slot from routine isotope production was in good agreement with observed activation products from monitor reactions. Spectral adjustment by maximizing the cross entropy further improved agreement between simulation and experiment by 8% on average for all reactions and 22% for $^{209}$Bi(n,6n)$^{204}$Bi. In comparison to past work at IPF, the activities produced in this work were generally >2x higher per incident proton and target mass, potentially due to the higher incident proton energy. For non-monitor reactions, major deviations in residual activity may be due to a lack of nuclear data and/or poor isomeric generation in FLUKA. Additionally, the generation of secondary protons/α-particles may confound analysis though directly simulated activities in FLUKA should take these into consideration. Simulations showed decreasing the degrader distance from the N-slot led to improvements in the fast secondary neutron yield, where W was identified as an ideal degrader candidate based on cost and its physical and thermal characteristics. From folding the FLUKA predicted neutron flux with TENDL cross sections, the best performing target array could produce a mass normalized saturation yield of ~0.6 mCi/µA-g $^{225}$Ra. The generation of other radionuclides using the N-slot could be useful for preclinical and research applications.

# 6 Acknowledgements

Brookhaven National Laboratory (BNL) is operated by Brookhaven Science Associates, LLC, (BSA) under a contract with the US Department of Energy (DOE, Contract No. DE-SC0012704). This research is supported by the U.S. Department of Energy Isotope Program, managed by the Office of Science for Isotope R&D and Production. WL acknowledges the support of the Natural Sciences and Engineering Research Council of Canada (NSERC), NSERC-PGSD-578011-2023, when developing the spectral adjustment code during graduate studies at the University of Wisconsin-Madison. The authors acknowledge all members of the BNL Medical Isotope Research and Production group for their assistance. We are grateful to BNL Radiological Control Division for the Health Physics support. Sumanta Nayak is acknowledged for assistance in engineering design of holders and degraders. We thank Dr. Jonathan W. Engle from University of Wisconsin-Madison for helpful discussions on spectral adjustment, Dr. Dohyun Kim for discussions related to leakage/secondary protons and Dr. Mary PW Chin for resolving logistical challenges regarding FLUKA licensing.

**Contributions**

**Wilson Lin:** Conceptualization, Data curation, Formal analysis, Investigation, Methodology, Software, Validation, Visualization, Writing-Original draft, Writing-Reviewing and Editing **Michael A. Skulski:** Conceptualization, Methodology, Investigation, Writing-Reviewing and Editing **Cathy S. Cutler:** Resources, Writing-Reviewing and Editing, Funding acquisition **Dmitri G. Medvedev:** Conceptualization, Methodology, Resources, Writing-Reviewing and Editing, Supervision, Project administration, Funding acquisition **Jonathan T. Morrell:** Conceptualization, Methodology, Writing-Reviewing and Editing, Supervision, Project administration